\newif\ifoverleaf
\ifoverleaf
\documentclass[twocolumn]{article}
\usepackage[margin=1in]{geometry}
\else
\documentclass[10pt,conference]{IEEEtran}
\fi

\usepackage{amsmath,amssymb,amsfonts}
\usepackage{graphicx}
\usepackage{hyperref}

\begin{document}

%%%%%%%%%%%%%%%%%%%%%%%%%%%%%%%%%%%%%%%%%%%%%%%%%%%%%%%%%%

\title{An Adversarial Quantum Key Distribution Project}

\ifoverleaf
\author{Brian R. La Cour and Noah A. Davis}
\else
\author{\IEEEauthorblockN{Brian R. La Cour}
\IEEEauthorblockA{\textit{Applied Research Laboratories} \\
\textit{The University of Texas at Austin}\\
Austin, Texas, United States \\
blacour@arlut.utexas.edu}
\and
\IEEEauthorblockN{Noah A. Davis}
\IEEEauthorblockA{\textit{Applied Research Laboratories} \\
\textit{The University of Texas at Austin}\\
Austin, Texas, United States \\
noah.davis@arlut.utexas.edu}
}
\fi

\maketitle

\begin{abstract}
Quantum key distribution (QKD) is a popular introduction to quantum technologies used in education and public outreach, as very little background in quantum theory is needed and the practical applications are easily understood.  There is considerably less exposure to the many real-world considerations of practical QKD, as access to the necessary hardware is quite limited.  Here we describe a simple, simulation-based QKD project that can be implemented with only a minimal background in quantum concepts and programming.  Students are assembled in small groups to develop an ``Alice and Bob'' protocol for securely distributing symmetric keys in a simulated noisy channel.  Their protocol is then shared anonymously with another group who plays the role of Eve and attempts to steal as much secret key as possible.  The adversarial aspect is popular with students, and the project itself provides a deeper understanding and appreciation for practical QKD.
\end{abstract}

% keywords: quantum key distribution, quantum education, quantum optics

%%%%%%%%%%%%%%%%%%%%%%%%%%%%%%%%%%%%%%%%%%%%%%%%%%%%%%%%%%

\section{Introduction}

Quantum mechanics poses many conceptual challenges, even for experts, and this makes effective pedagogical methods in teaching quantum concepts at the introductory level particularly difficult \cite{Singh2001,Muller2002}.  In recent years, this problem has come to the fore with the increased interest in quantum information science and the various attempts to introduce it at the undergraduate, high school, and even middle school level \cite{Weissman2024,Zuccarini2024}.  A popular approach to introducing quantum concepts is through an introduction to quantum key distribution (QKD) methods, in particular the 1984 protocol of Bennett and Brassard (BB84) \cite{BB84,Svozil2006,Kohnle2017,DeVore2020,Utama2020}.

The BB84 protocol is appealing as a pedagogical tool for a variety of reasons.  It requires very little background in quantum physics, yet covers most of the essential aspects of quantum information science (QIS), such as superposition and quantum measurement.  Its description in terms of light polarization is easily grasped from analogous and familiar concepts in classical transverse waves.  Even the concept of discrete photons can be conveniently, albeit incorrectly, illustrated as pulses of classical light.  It is only in quantum measurement that a true departure from classical physics is needed, and even that can be scaffolded with a discussion of Malus's law.  Furthermore, QKD provides an excellent vehicle to introduce other central concepts, such as the no-cloning theorem \cite{Wootters1982}.  Of course, the practical aspect of QKD as a tool for secure communication is perhaps its greatest appeal, and this provides a useful context to introduce basic concepts in cryptography, such as public and symmetric key encryption.

We have used QKD as an early introduction to QIS applications at both the high school and early undergraduate level \cite{Walsh2021,Davis2022}.  In 2018 we started the first year-long program in QIS education for first-year undergraduates.  Starting each spring, students are introduced to basic concepts in QIS through a description of analogous classical optics phenomena, focusing on applications to quantum communication \cite{Gisin2002}.  QKD is typically introduced in the third week and is the focus of a midterm project that has small groups of students working in a fun yet adversarial effort to eavesdrop on one another!  Games such as this have proven a very effective pedagogical tool in teaching QIS concepts \cite{Seskir2022}.

The project is implemented in four parts, following an initial introduction to QKD and the BB84 protocol, and students work in small assigned groups.  In Part 1, students are given a Jupyter notebook containing code stubs that are completed to implement an ideal BB84 QKD protocol.  Part 2 uses a more sophisticated simulation model that incorporates various nonidealities, such as photon loss and dark counts.  Each group implements an ``Alice and Bob'' protocol that attempts to detect a potential eavesdropper in the presence of environmental noise while maximizing their secure key rate.  In Part 3, each group is given a different group's ``Alice and Bob'' protocol and tasked with developing an ``Eve'' protocol that maximizes the stolen key rate while minimizing the probability that they are detected.  Each group writes a concise, roughly two-page report and prepares a short, four-minute presentation on their work.  Presentations are given in a ``mini-symposium'' format that encourages students to actively engage with their peers during a one-minute Q\&A session following each talk.  Note that, because this project prioritizes accessibility to novices in the field, the code stubs provided to students and the examples in this work are written to emphasize clarity rather than efficiency.

The organization of the paper is as follows: In Sec.\ \ref{sec:IdealQKD} we describe Part 1 of the project, wherein individual students implement an ideal BB84 protocol using the Qubit class in Python.  In Sec.\ \ref{sec:PracticalQKD} we describe the Photon class implemented in Python to represent non-ideal, practical aspects of realistic QKD system.  Section \ref{sec:AliceBob} discusses the ``Alice and Bob'' protocol, while Sec.\ \ref{sec:Eve} discusses that of Eve, now in an adversarial role.  Metrics for assessing the performance of each protocol are described in Sec.\ \ref{sec:Assessment}, and some common pitfalls in students' implementations are surveyed in Sec.\ \ref{sec:Missteps}.  Our conclusions are summarized in Sec.\ \ref{sec:Conclusions}.  Project details, instructional materials, and the notebooks provided to students are available through Canvas Commons \cite{Canvas}.

%%%%%%%%%%%%%%%%%%%%%%%%%%%%%%%%%%%%%%%%%%%%%%%%%%%%%%%%%%

\section{Ideal QKD with Qubits}
\label{sec:IdealQKD}

The code stubs provided to students in Part 1 of the project are given in the form a Jupyter notebook containing Python code.  Such notebooks can be accessed and run online through services such as Google Colaboratory, MyBinder, CoCalc, JupyterLite, etc.  Early versions of this project used stand-alone Python scripts that required students to install their own Python development environment.  This proved a practical difficulty, as students work with a variety of hardware and operating system baselines.  In a high school environment, with its many restrictions on installing software, it was not even an option.  Online access is therefore highly desirable and encouraged.  Notebooks are further encouraged, as they provide a means of modularly organizing the code and allow one to intersperse text, figures, and even equations through the use of MarkDown language.

The notebook implements a Qubit class, with basic methods to prepare qubits in one of several polarization states, such as horizontal (H), vertical (V), diagonal (D), anti-diagonal (A), right-circular (R), left-circular (L), and measure them in different bases (i.e., H/V, D/A, R/L).  The use of polarization provides a simple visualization of the different qubit states and a touchpoint to real-world QKD systems (at least, those operating in free space, which tend to be polarization based).  Measurement outcomes are probabilistic, strictly follow the Born rule, and offer no invalid or missing results.  Notebooks allow one to collapse the cell containing the Qubit class for greater readability.  Students can see the code but can only interact with the qubits using the prescribed methods, thereby providing a crude representation of their elusive quantum nature.  Besides the Qubit class, the only required packages are \texttt{random} and \texttt{numpy}.

The four remaining notebook cells are Alice, Eve, Bob, and Key Sifting.  The Alice cell defines the number of qubits, $n$, to be used.  To minimize the use of exotic data structures, strings of length $n$ are used to represent most objects.  For example, Alice begins by defining a raw key as a length-$n$ string of `0' and `1' characters.  The following sample code is provided to the students:
\begin{verbatim}
# Alice generates the raw key.
keyAlice = "" # Initialize the string.
# Iterate over the number of qubits.
for i in range(n):
  # Append a random character 
  # ('0' or '1') to the end.
  keyAlice += random.choice(['0','1'])
print("keyAlice    = " + keyAlice)
\end{verbatim}

With this template, students may implement other parts of the protocol.  For example, Alice's next task is to choose a random basis to represent each bit of the key.  This, too, is represented by a string of length $n$, with `+' representing the horizontal/vertical (H/V) basis and `x' representing the diagonal/anti-diagonal (D/A) basis.  For scaffolding, students are provided partial code, with a ``\texttt{\# TODO: Put your code here.}'' comment added where the student is expected to add their contribution.  For example, the next step for Alice is given as follows:
\begin{verbatim}
# Alice chooses the encoding basis for 
# each key bit.  This should be a string 
# of '+'s and 'x's with '+'=H/V, 'x'=D/A.
basisAlice = ""
# TODO: Put your code here.
print("basisAlice  = " + basisAlice)
\end{verbatim}
In place of the ``\texttt{TODO}'' line the student is then expected to add the following lines:
\begin{verbatim}
for i in range(n):
  basisAlice += random.choice(['+','x'])
\end{verbatim}

This scaffolding approach allows students with minimal programming experience to build off of prior templates and generate increasingly sophisticated code elements.  The next step adds the use of conditionals, as Alice selects which polarization to use for each element of the raw key.  To do this, students build up a string, \texttt{qubitAlice}, of length $n$ consisting of the four characters `H', `V', `D', and `A' by combining a for-loop with a set of nested conditionals.  One example of this code snippet is the following:
\begin{verbatim}
qubitAlice = ""
# TODO: Put your code here.
for i in range(n):
  if basisAlice[i]=='+':
    if keyAlice[i]=='0':
      qubitAlice += 'H'
    elif keyAlice[i]=='1':
      qubitAlice += 'V'
  elif basisAlice[i]=='x':
    if keyAlice[i]=='0':
      qubitAlice += 'D'
    elif keyAlice[i]=='1':
      qubitAlice += 'A'
print("qubitAlice  = " + qubitAlice)
\end{verbatim}

Finally, students use the Qubit class to generate an array of Qubit objects.  This is perhaps the most exotic data structure they will need to work with, and, again, scaffolding is provided as an aid.  The students are given the following:
\begin{verbatim}
qubitArray 
= [qubit.Qubit() for i in range(n)]
# TODO: Put your code here.
for i in range(n):
  if   qubitAlice[i]=='H':
      qubitArray[i].prepareH()
  elif qubitAlice[i]=='V':
      qubitArray[i].prepareV()
  elif qubitAlice[i]=='D':
      qubitArray[i].prepareD()
  elif qubitAlice[i]=='A':
      qubitArray[i].prepareA()
\end{verbatim}

There is a cell for Eve that students are instructed to leave alone until they have completed their Alice and Bob protocols.  It is essentially just an implementation of Bob's protocol, followed by an implementation of Alice's protocol in a standard intercept-and-resend attack.

Code for the Bob cell follows a similar structure to that of Alice.  Bob begins by constructing a string of length $n$, \texttt{basisBob}, indicating the bases in which he randomly chooses to meaure.  This is followed by construction of an \texttt{outcomeBob} string of the four polarization characters, which are generated by the \texttt{measureHV} and \texttt{measureDA} methods within the Qubit class definition.  Finally, Bob infers his own raw key, \texttt{keyBob}, as a string of `0' and `1' characters depending upon the outcome of each measurement.  Statistically, only about half of Bob's raw key should match that of Alice.

The final cell performs key sifting.  The students implement code to build two strings, \texttt{siftedAlice} and \texttt{siftedBob}, each of length $n$, that contain the sifted key.  The students are instructed to place a delimiter, `-`, in places where the two bases do not match.  Otherwise, Alice and Bob place the elements of their respective raw keys.  Using a delimiter ensures that all strings are kept at length $n$ and facilitates both comparison and debugging.  A final code snippet is provided to the students to analyze their results.
\begin{verbatim}
# Compare Alice and Bob's sifted keys.
numMatch = 0
for i in range(len(siftedAlice)):
  if siftedAlice[i] == siftedBob[i]:
    numMatch += 1
matchPercent 
= numMatch / len(siftedAlice) * 100
print(str(matchPercent) + "% match")
\end{verbatim}

Once the students have verified that their protocol, without Eve, yields a perfect match with a sifted key of about $n/2$ bits, they can implement the Eve cell and verify that the match drops to about 75\%, indicating the presence of an eavesdropper.

%%%%%%%%%%%%%%%%%%%%%%%%%%%%%%%%%%%%%%%%%%%%%%%%%%%%%%%%%%

\section{Practical QKD with Photons}
\label{sec:PracticalQKD}

The Ideal QKD notebook is designed to be completed by each individual student and is intended to both bring them up to speed on basic programming and to reinforce the basic elements of the BB84 protocol.  The next part of the project has students work in groups of two-to-three individuals to reimplement the Alice-and-Bob protocol in the presence of channel noise and loss.  This is done through the introduction of the Photon class, a generalization of the Qubit class used by the students earlier.

The Photon class is based on dual-mode weak coherent light as an approximate source of single photons and is an early predecessor to our online Virtual Quantum Optics Laboratory \cite{VQOL}.  Internally, this is modeled using a pair of complex Gaussian random variables, $a_H$ and $a_V$, representing the horizontal and vertical components, respectively, of the polarization Jones vector.  If $|{\alpha_H}\rangle \otimes |{\alpha_V}\rangle$ is the dual mode weak coherent state, where $\alpha_H, \alpha_V \in \mathbb{C}$ are the complex amplitudes, then
\begin{subequations}
\begin{align}
a_H &= \alpha_H + \sigma_0 z_H \\
a_V &= \alpha_V + \sigma_0 z_V \; ,
\end{align}
\end{subequations}
where $\sigma_0 = 1/\sqrt{2}$ and $z_H$, $z_V$ are independent standard complex Gaussian random variables with zero mean and unit variance.  This representation follows from the Wigner function of the quantum state, which is Gaussian in nature.  The Photon class provides a \texttt{prepare} method that takes as its arguments the variables $\psi_H$, $\psi_V$, and $\mu$ representing the two complex components of the qubit state, where $|\psi_H|^2 + |\psi_V|^2 = 1$, and the average photon number $\mu \ge 0$.  The coherent state parameters are then
\begin{subequations}
\begin{align}
\alpha_H &= \sqrt{\mu} \, \psi_H \\
\alpha_V &= \sqrt{\mu} \, \psi_V \; ,
\end{align}
\end{subequations}
Note that if $\mu = 0$ then $a_H$ and $a_V$ represent vacuum states.

Similar to the Qubit class, the Photon class offers a measurement method, \texttt{measureHV}, that performs measurements in the H/V basis.  (Similar methods exist for the D/A and R/L bases.)  This method differs from that of the Qubit class in providing \emph{four} possible outcomes: `H', `V', `N', and `M', where `H' and `V' are considered valid outcomes of either horizontal or vertical polarization, while `N' and `M' are invalid outcomes of either no detections or multiple (dual) detections, respectively.  Physically, it is modeled as a polarizing beam splitter with photon detectors at each output port.  The method is defined mathematically as follows:
\begin{equation}
(a_H, a_V, p_d) \mapsto \begin{cases}
\text{`H'} & \text{if $|a_H| > \gamma$ and $|a_V| \le \gamma$,} \\
\text{`V'} & \text{if $|a_H| \le \gamma$ and $|a_V| > \gamma$,} \\
\text{`N'} & \text{if $|a_H| \le \gamma$ and $|a_V| \le \gamma$,} \\
\text{`M'} & \text{otherwise}
\end{cases}
\label{eqn:measurementOutcomes}
\end{equation}
where $p_d \in [0,1]$ is the probability of a dark count and $\gamma \ge 0$ is the corresponding detection threshold, given by
\begin{equation}
\gamma^2 = -\sigma_0^2 \log\left( 1 - \sqrt{1 - p_d} \right) \; .
\label{eqn:threshold}
\end{equation}
The parameter $p_d$ is selected by the students.  Small values of $p_d$ lead to many `N' outcomes, while large values lead to many `M' outcomes.  Thus, there is a tradeoff that is part of the design of the students' protocol, and they should be encouraged to explore that tradeoff.

Several other linear optical components are available as methods within the Photon class.  They include \texttt{applyUnitaryGate}, which applies a general three-parameter single-qubit unitary, along with several specific instances such as Pauli and Hadamard gates.  Other methods implement non-unitary operations, such as \texttt{applyPolarizer}, which projects onto a given polarization (and produces an complementary vacuum state), and \texttt{applyAttenuation}, which acts as a neutral density filter.  Finally, there is a depolarizing method, \texttt{applyNoisyGate}, which applies a Haar-random unitary.  These various operations are not required for implementing the BB84 protocol, but can serve as a ``bag of tricks'' for an eavesdropper wishing to cover their tracks.

%%%%%%%%%%%%%%%%%%%%%%%%%%%%%%%%%%%%%%%%%%%%%%%%%%%%%%%%%%

\section{Alice and Bob's Protocol}
\label{sec:AliceBob}

The students are once again provided with a notebook containing program scaffolding and expected to write their own code in sections marked \texttt{TODO: Put your code here.} The basic structure parallels that of the ideal implementation with the \texttt{Qubit} class replaced by the \texttt{Photon} class. For example, once Alice's bases and key bits are chosen, the photon array might be prepared as follows with an average photon number of, say, $\mu = 5$:
\begin{verbatim}
photonArray = [Photon() for i in range(n)]
# TODO: Put your code here.
avgPhotNum = 5 # Average photon number
for i in range(n):
  if photonAlice[i] == 'H':
    photonArray[i].prepareH(avgPhotNum)
  elif photonAlice[i] == 'V':
    photonArray[i].prepareV(avgPhotNum)
  elif photonAlice[i] == 'D':
    photonArray[i].prepareD(avgPhotNum)
  elif photonAlice[i] == 'A':
    photonArray[i].prepareA(avgPhotNum)
\end{verbatim}

A possible point of confusion for students is that the average photon number represents a continuously variable amplitude for the coherent state and \emph{not} a discrete number of photons.  This should be contrasted with the variable $n$, which defines the number of elements in the photon array and physically represents the number of coherence times within the duration of Alice's transmission.

As in the ideal case, Bob generates a string of basis choices and measures the photon array using either the \texttt{measureHV} or \texttt{measureDA} method. Unlike the ideal case, invalid measurements are now possible, as detailed in Eqn.\ (\ref{eqn:measurementOutcomes}). In fact, the Photon class measurement methods include a parameter, $p_d$, specifying the dark count probability and corresponding detection threshold, as described in Eqn.\ (\ref{eqn:threshold}). Students are encouraged to experiment with this parameter and tune their detectors as they see fit.  When Bob infers the bit values of the raw key from the measurement outcomes, he places a dash (`-') wherever a measurement is invalid.  This serves to keep Bob's raw key the same length as Alice's and facilitates later analysis.

The final communication step in the BB84 protocol, before any analysis of security, is for Alice and Bob to sift their keys. The public revelation of basis choice is accomplished by simply allowing access to the basis arrays. In this project, Bob's invalid measurements are also publicly revealed. The students are expected to construct sifted keys by checking if Alice and Bob have chosen the same basis for each key bit. For example,
\begin{verbatim}
# Alice and Bob extract their sifted keys.
# siftedAlice and siftedBob should 
#   be strings of length n.
# Use the convention '0', '1', ' '=removed
siftedAlice = ""
siftedBob   = ""
# TODO: Put your code here.
for i in range(n):
  if keyBob[i] == '-':
    siftedAlice += ' '
    siftedBob += ' '
  elif basisAlice[i] == basisBob[i]:
    siftedAlice += keyAlice[i]
    siftedBob += keyBob[i]
  else:
    siftedAlice += ' '
    siftedBob += ' '
\end{verbatim}
At this point, Alice and Bob have keys of length $n$ containing blank spaces at invalid or mismatched measurements and their respective secret keys where the valid measurement bases match.

Unlike in the Ideal QKD case, the secret keys of Alice and Bob may not match, even in the absence of an eavesdropper.  This is due to the various nonidealities within the Photon class and may physically be attributed to decoherence and photon loss within the channel.  As will be described later, students must assess and characterize this baseline level of system and environmental noise in order to detect an eavesdropper whose presence will, presumably, increase the frequency of errors by a noticeable degree.

%%%%%%%%%%%%%%%%%%%%%%%%%%%%%%%%%%%%%%%%%%%%%%%%%%%%%%%%%%

\section{Eve Attacks!}
\label{sec:Eve}

As in the ideal case, students are asked not to complete the Eve portions of the code while working on their Alice and Bob implementations.  (Some may choose to do so in order to ``stress test'' their protocol, but the final code should have Eve removed.)  Unlike in the idealized portion of the project, the student groups are matched adversarially; each group's Alice and Bob implementation is anonymized and given to another, random group to act as an attacking eavesdropper, Eve.  Eve traditionally has the goal of stealing as much secret key as possible without being detected. The groups have full read-access to their adversary's Alice-and-Bob code but may not use unphysical techniques, such as directly looking at Alice's chosen bits or bases.

Eve's code is interleaved with Alice and Bob's in a single notebook. The first section asks Eve to select some sub-sample of photon array elements to measure by specifying indices of \texttt{photonArray} as a binary string, \texttt{sampleIndex}, with values of 1 indicating array elements to be sampled. Eve is free to choose as many or as few samples as desired and applies the chosen measurements with her own choice of dark count probability.  (Eve may, after all, own a different type of detector than Bob.)  The eavesdropper may then prepare a replacement signal for any photons lost to measurement. Eve is encouraged at this point to add any other nasty tricks to take advantage of vulnerabilities discovered in the Alice-Bob protocol.  For example, she may choose to simply add noise to some of the photons rather than measure them in order to alter Alice and Bob's assessment of the baseline channel noise.

Eve's final section comes after Alice and Bob have sifted their keys. Using the publicly revealed information, the Eve group can sift their own stolen key and choose which bits to keep. The eavesdroppers are expected to throw out any bits in which all three parties' chosen bases do not match or if measurement results were invalid. Eve's key sifting might look like the following:
\begin{verbatim}
# stolenEve should be strings of length n.
# Use the '0', '1', ' '=removed
stolenEve = ""
# TODO: Put your code here.
for i in range(n):
  if sampleIndex[i] == '0':
    stolenEve += ' '
  elif basisAlice[i] != basisBob[i]:
    stolenEve += ' '
  elif basisAlice[i] != basisEve[i]:
    stolenEve += ' '
  elif keyBob[i] == '-':
    stolenEve += ' '
  elif keyEve[i] == '-':
    stolenEve += ' '
  else:
    stolenEve += keyEve[i]
\end{verbatim}
By the end of communication, Eve has stolen some amount of secret key. Alice and Bob then go about the process of attempting to evaluate the security and effectiveness of the key sharing protocol.

A variant of the project, which we have used in high school settings, uses a set of predefined Eve protocols as adversaries, with each group assigned a randomly chosen, predefined Eve.  This approach can facilitate implementation at scale, where rubrics can be tailored to known eavesdropper techniques, but it lacks the peer-to-peer competitive nature that makes this project so appealing to many students.

%%%%%%%%%%%%%%%%%%%%%%%%%%%%%%%%%%%%%%%%%%%%%%%%%%%%%%%%%%

\section{Assessing Protocol Performance}
\label{sec:Assessment}

With quantum communication completed and keys sifted, Alice and Bob choose a subset of their sifted key to compare publicly in order to evaluate the security of the channel. Much like when Eve selected indices for the photons to intercept, Alice and Bob specify indices of bits to sample. They compare Alice's sent signal with Bob's measured signal and keep track of the number of bits that do not match. This is then used to calculate the sample quantum bit error rate (QBER) as a diagnostic metric. Below is an example of sampling the QBER while checking approximately one fifth of the bits:
\begin{verbatim}
sampleIndex = ""
sampledBobQBER = 0
# TODO: Put your code here.
mismatch = 0
totalSample = 0
for i in range(n):
  if random.random() < 0.2:
    if siftedAlice[i] != ' ':
      sampleIndex += '1'
      totalSample += 1
      if siftedAlice[i] != siftedBob[i]:
        mismatch += 1
    else:
      sampleIndex += '0'
  else:
    sampleIndex += '0'
if (totalSample != 0):
  sampledBobQBER = mismatch/totalSample
\end{verbatim}
Alice and Bob then use the sampled QBER to determine whether their channel is secure. Students are encouraged to try a few strategies for choosing bits and determining security to balance their risk tolerance against their communication rate.

The final analysis is included in the notebook and requires no input from the students directly; however, they may use it for inspiration and to evaluate their protocol while in development. Alice and Bob's actual quantum bit error rate (QBER) is calculated by comparing their secure keys, and the secure key rate is given based on the QBER, the number of bits initially sent, and the final length of the secure key. Eve's QBER and the stolen key rate are similarly calculated and all the results are printed to the screen (which makes a nice visual for the student presentations).
\begin{verbatim}
if not channelSecure:
  secureKeyRateBob = 0;
  stolenKeyRateEve = 0;
  print("***********************")
  print("* ALERT! *")
  print("Quantum channel not secure")
  print("* ALERT! *")
  print("***********************")
\end{verbatim}

As a further assessment of performance, it can be instructive to have students use their protocol to actually send and receive encoded messages.  Having students design a full, end-to-end encoding and decoding scheme can be instructive for understanding how QKD systems are actually used for secure communication.

%%%%%%%%%%%%%%%%%%%%%%%%%%%%%%%%%%%%%%%%%%%%%%%%%%%%%%%%%%

\section{Common Missteps}
\label{sec:Missteps}

Students on both the sending/receiving side and the eavesdropping side of the protocol employ a number of interesting strategies that sometimes become vulnerabilities.  Our experience is that, by and large, students are clever enough to identify vulnerabilities in their adversary's protocol yet careless (or overly clever) enough to create vulnerabilities of their own.  Instructors are advised not to give too much advice on how best to design their protocols.  The missteps are a valuable part of the learning process!

One popular attempt by the Alice-and-Bob team is to obfuscate their protocol by changing variable names, interchanging the H/V basis with the D/A basis, mapping 0 to 1, or other, similar bids. These almost universally fail to deceive the eavesdropper, because the students acting as Eve have read-access to the Alice-and-Bob code and can study their protocol in detail. Similarly, Alice will sometimes set a classical password for Bob to use when choosing measurement bases; since the password is set in the code, Eve has direct access to it.

The sampling of check-bits in the sifted key is another common step that introduces vulnerabilities. Some Alice-and-Bob protocols do not sample the key at all or, worse, sample the entire key!  This allows either complete, unmonitored access of the key to Eve or publishes all existing bits for an eavesdropper to harvest at their leisure. The choice of key fraction to sample presents a delicate balance that must be struck between security and key rate. More commonly, Alice and Bob will compare a reasonable fraction of the key bits but will always sample the same portion---for example, the first 10\%. Once the Eve group notices this vulnerability in the protocol, they have free reign to steal all remaining key bits without detection.

Beyond the basic protocol, students have the opportunity to tune a number of parameters related to the practical implementation of QKD, including the number of bits to send, the power of each pulse (i.e., the average photon number), and the probability of dark counts (which controls detector sensitivity). Sending too few bits allows for wildly inconsistent statistics and can hide other problems; students usually realize this while experimenting but often take $n$ to be far too small for adequate characterization. An average photon number, $\mu$, set too low or too high can result in abundant nondetections or multidetections respectively, significantly reducing their sample size. More subtly, a low or high average photon number can mask the presence of an eavesdropper by either allowing incorrectly retransmitted bits to blend into the expected nondetections or by allowing the eavesdropper to make multiple measurements in a beamsplitter attack.

Most mistakes on Eve's part come from the idea that denial of service is a desired outcome. Eavesdroppers tend to either measure too often or add too much noise and accept that the channel will be flagged as insecure.

%%%%%%%%%%%%%%%%%%%%%%%%%%%%%%%%%%%%%%%%%%%%%%%%%%%%%%%%%%

\section{Conclusions}
\label{sec:Conclusions}

We have presented a group-based project in which students with no prior knowledge of programming and minimal exposure to quantum concepts are guided through an implementation of the BB84 quantum key distribution protocol. Jupyter notebooks containing Python code stubs and hints help those new to the topic of quantum communication and lower the barrier to entry for novice programmers. Having the students first work through an idealized version of the protocol individually helps them better tackle the considerations of practical optics as they work in a group setting.  

Although many demonstrations of QKD emphasize the ideal nature of the protocol, we have found that the introduction of practical considerations, while a potential distraction, actually offers a much richer context in which students can exercise their creativity and problem solving abilities.   Students are given a wide array of tools with which to implement their protocols, and we have found that they use them in suprising and novel ways.  We believe the open-ended nature of the project is a key strength and one that students find engaging.  In addition, the adversarial nature of the project adds an element of excitement and competition that, we have found, provides students with a strong sense of motivation.

%%%%%%%%%%%%%%%%%%%%%%%%%%%%%%%%%%%%%%%%%%%%%%%%%%%%%%%%%%

\section*{Acknowledgments}

This work was supported by the National Science Foundation (NSF) under Grant No. 1842086 and by the Freshman Research Initiative (FRI) program under the College of Natural Sciences at The University of Texas at Austin.

%%%%%%%%%%%%%%%%%%%%%%%%%%%%%%%%%%%%%%%%%%%%%%%%%%%%%%%%%%%
%\ifoverleaf
%\bibliographystyle{unsrt}
%\else
%\bibliographystyle{IEEEtran}
%\fi
%\bibliography{refs}

% Generated by IEEEtran.bst, version: 1.14 (2015/08/26)

\end{document}